\documentclass[11pt]{article}
\usepackage{epsfig}

\newcommand{\beq}{\begin{equation}}
\newcommand{\eeq}{\end{equation}}
\newcommand{\beqa}{\begin{eqnarray}}
\newcommand{\eeqa}{\end{eqnarray}}

\newcommand{\laem}{\stackrel{<}{\sim}}

\begin{document}

\begin{titlepage}
\def\thepage {}        

\title{Saturating the Bound on the Scale of Fermion Mass Generation}

\author{
R. Sekhar Chivukula\thanks{e-mail addresses: sekhar@bu.edu}\\
Department of Physics, Boston University, \\
590 Commonwealth Ave., Boston MA  02215}

\date{July 16, 1998}

\maketitle

\bigskip
\begin{picture}(0,0)(0,0)
\put(295,250){BUHEP-98-21}
\put(295,235){hep-ph/9807406}
\end{picture}
\vspace{24pt}

\begin{abstract}

  Recently, J\"ager and Willenbrock have shown that the Appelquist and
  Chanowitz bound on the scale of top-quark mass generation can formally
  be saturated at tree-level in a particular limit of a two-Higgs
  doublet model.  In this note I present an alternate derivation of
  their result. I perform a coupled channel analysis for $f\bar{f} \to
  V_L V_L$ and $V_L V_L \to V_L V_L$ scattering and derive the
  conditions on the parameters required for $f \bar{f} \to V_L V_L$
  scattering to be relevant to unitarity. I also show that it is not
  possible to saturate the bound on fermion mass generation in a
  two-Higgs model while maintaining tree-level unitarity in Higgs
  scattering at high energies.

\pagestyle{empty}
\end{abstract}
\end{titlepage}

\setcounter{section}{1}

\setcounter{equation}{0}

Appelquist and Chanowitz \cite{Appelquist:1987cf} derived an upper bound
on the scale of fermion mass generation by examining the inelastic
scattering amplitude $f \bar{f} \to V_L V_L$, where $V_L$ denotes
longitudinally polarized $W$ or $Z$ gauge bosons.  In the absence of the
Higgs boson, or another dynamics responsible for generating fermion
mass, this amplitude grows with increasing center-of-mass energy. This
tree-level amplitude would ultimately violate unitarity at a
sufficiently high energy $\Lambda_f$.  Therefore, one concludes that the
scale associated with fermion mass generation is bounded by $\Lambda_f$.
The strictest bound for a fermion with mass $m_f$ is obtained from the
spin-zero, weak-isosinglet, color-singlet amplitude
\cite{Marciano:1989ns}
\beq
\Lambda_f < {8\pi v^2 \over \sqrt{3N_c} m_f}~,
\label{applchanbound}
\eeq
where $v=246$ GeV and $N_c$ is the number of colors (3 for quarks and 1
for leptons).
The strongest bound occurs for the top quark. With $m_t \approx 175$ GeV,
the upper bound (\ref{applchanbound}) on the scale of top mass generation
is $\Lambda_t\approx 3$ TeV. 

On the other hand, in the absence of a Higgs boson or some other
dynamics responsible for electroweak gauge-boson mass generation, the
elastic scattering amplitude $V_L V_L \to V_L V_L$ grows quadratically
with center-of-mass energy. Therefore, one concludes that the scale
associated with gauge boson mass generation is bounded by a scale
$\Lambda_{EW}$ where the elastic scattering amplitude would violate
unitarity. The strictest bound is obtained from the spin-zero,
weak-isosinglet scattering amplitude
\cite{Lee:1977eg,Chanowitz:1985hj,Marciano:1989ns}
\beq
\Lambda_{EW} < \sqrt{8\pi} v \approx 1.2\, {\rm TeV}~.
\eeq
Since $\Lambda_{EW}$ is less than $\Lambda_t$, it is not clear that the
bound (\ref{applchanbound}) is relevant
\cite{Appelquist:1987cf,Golden:1994pj}. The physics responsible
for unitarizing the elastic gauge boson scattering amplitude may
unitarize the inelastic $t\bar{t} \to V_L V_L$ amplitude, as happens in
the standard one-doublet Higgs model.

Recently, J\"ager and Willenbrock \cite{Jaeger:1998va} have shown that
the Appelquist and Chanowitz \cite{Appelquist:1987cf} bound
(\ref{applchanbound}) on the scale of top quark mass generation can
formally be saturated at tree-level in a particular limit of a two-Higgs
doublet model.  In this note I present an alternate derivation of their
result. I perform a coupled channel analysis for $f\bar{f} \to V_L
V_L$ and $V_L V_L \to V_L V_L$ scattering and derive the conditions on
the parameters required for $f \bar{f} \to V_L V_L$ scattering to be
relevant to unitarity. I also show that it is not possible to saturate
the bound on fermion mass generation in a two-Higgs model while
maintaining tree-level unitarity in Higgs scattering at high energies.

Consider the general potential in a two-Higgs model written in the
form \cite{Georgi:1978xz}
\beqa
\hskip-5pt V(\phi_1,\phi_2) &=&  \lambda_1(\phi^\dagger_1\phi_1-{ v_1^2}/2)^2
 + \lambda_2(\phi^\dagger_2\phi_2-{ v_2^2}/2)^2 \nonumber \\
&+&\lambda_3\left[(\phi^\dagger_1\phi_1-{ v_1^2}/2) 
 + (\phi^\dagger_2\phi_2-{ v_2^2}/2)\right]^2 \nonumber
 \\
&+& { \lambda_4}\left[(\phi^\dagger_1\phi_1)(\phi^\dagger_2\phi_2)
- (\phi^\dagger_1\phi_2)(\phi^\dagger_2\phi_1)\right] \nonumber \\
&+& { \lambda_5}\left[{\rm Re}(\phi^\dagger_1\phi_2) 
       - { v_1}{ v_2}\cos\xi/2\right]^2 \nonumber \\
&+& { \lambda_6}\left[{\rm Im}(\phi^\dagger_1\phi_2) 
       - { v_1}{ v_2}\sin\xi/2\right]^2  ~,
\label{eq:twohiggpot}
\eeqa
where $\phi_1$ and $\phi_2$ are weak doublet scalar fields with
hypercharge $+1/2$. For simplicity, in the following we will set
$\xi=0$.  In order to obtain the correct gauge boson masses, we must
require that
\beq
v^2_1 + v^2_2 = v^2 \approx (246\, {\rm GeV})^2~.
\eeq
We also impose a softly broken discrete symmetry under which $\phi_1$
and the right-handed down-quark and charged-lepton fields change sign.
This symmetry eliminates an extra terms which would otherwise have
been present in (\ref{eq:twohiggpot}) and it insures that only $\phi_2$
contributes to up-quark masses in general and the top-quark mass in
particular.

In this language, the limit considered by J\"ager and Willenbrock
\cite{Jaeger:1998va} corresponds to $\lambda_5 \to \infty$
\cite{Haber:1995be}, $\lambda_i$ (where $i=1-4$ or $6$) small, and $v_2
\ll v_1 \approx v$. Note that this is a non-decoupling limit
\cite{Golden:1994pj,Jaeger:1998va} in that a {\it dimensionless} coupling,
$\lambda_5$, is  taken large instead of a dimensionful one
\cite{Appelquist:1975tg}.

The vacuum expectation values (vevs) and neutral scalar fields can be
written
\beq
\phi_1 \to \left(\begin{array}{c} 0 \\ 
{{ (h_1+v_1)}/\sqrt{2}} \end{array}\right)
\ \ \ \ \&\ \ \ \  
\phi_2 \to \left(\begin{array}{c} 0 \\ 
{{(h_2+ v_2)} / \sqrt{2}} \end{array}\right)~.
\eeq
In the limit considered, the dominant contributions to the neutral
scalar masses come from the $\lambda_5$ term above, which gives:
\beq
{\lambda_5 \over 4} \left[ v_2 h_1 + v_1 h_2 + h_1 h_2 \right]^2~.
\label{lambdafiveterm}
\eeq
From this we immediately see that the combination
\beq
H={v_2 h_1 + v_1 h_2 \over \sqrt{v^2_1 + v^2_2}} 
\label{Hfield}
\eeq
has approximately the mass
\beq
m^2_{H} = {\lambda_5 v^2 \over 2}~.
\label{Hmass}
\eeq
In contrast, the orthogonal combination
\beq
h={- v_1 h_1 + v_2 h_2 \over \sqrt{v^2_1 + v^2_2}} 
\label{hfield}
\eeq
has mass
\beq
m^2_h = {\cal O}(\lambda_i v^2)
\eeq
with $i=1-4$ or 6, and remains light. In the two-Higgs notation employed
in \cite{Jaeger:1998va,Gunion:1989we}, these relations may be written
\beq
\cos\alpha \approx \sin\beta \approx {v_2\over v}~,
\label{mixone}
\eeq
and 
\beq
\sin\alpha \approx \cos\beta \approx {v_1\over v}~.
\label{mixtwo}
\eeq
In the limit $\lambda_5 \to \infty$ and $v_2 \ll v_1$, one can easily
verify that the corrections to the expressions given above for the
neutral scalar masses and mixings are suppressed by $\lambda_i /
\lambda_5$, $v_2/v$, or both.

Recalling that only $\phi_2$ couples to the top quark,
we find the couplings of the neutral scalar fields to the
top quark are
\beq
{\sqrt{2}m_t \over v}\, \bar{t}t\, \left( 
{\cos\alpha\over\sin\beta}\,h + {\sin\alpha\over \sin\beta}\, H\right)~.
\label{Httcoupling}
\eeq
Note that the coupling of $h$ to the top-quark is approximately equal to
that of the standard model Higgs, while the coupling of the $H$ is
enhanced by $v_1/v_2 \gg 1$.  The couplings of the neutral scalars to W
and Z gauge boson pairs is
\beq
{2\over v}
\left(\sin(\beta-\alpha)\, h + \cos(\beta-\alpha)\, H\right)
\left(M^2_W W^{\mu +} W_\mu^- + {1\over 2}M^2_Z
 Z^\mu Z_\mu \right)~.
\label{hvvcoupling}
\eeq
From eqns. (\ref{mixone}) and (\ref{mixtwo}) above, we
calculate
\beq
\sin(\beta-\alpha)\approx{{v^2_2 - v^2_1}\over v^2}
=\, -1 + {\cal O}\left({v^2_2\over v^2}\right)~,
\label{sinbetaalpha}
\eeq
and
\beq
\cos(\beta - \alpha) \approx {2v_1 v_2\over v^2}
= {2v_2\over v} + {\cal O}\left({v^3_2\over v^3}\right)~.
\label{cosbetaalpha}
\eeq

Next consider scattering amplitudes in the
range of energies $m_{W,Z}, m_h, m_t \ll \sqrt{s} \ll
m_H$. $V_L V_L$ elastic scattering is largely unitarized
by $h$ exchange. The properly normalized tree-level, spin-zero,
weak-isosinglet amplitude is
\beq
{\cal A}_{el}(V_LV_L\to V_LV_L)\vert_{\sqrt{s}\ll m_H}
= {s\over 16 \pi v^2}
\left( 1 - \sin^2(\beta-\alpha)\right) \approx {s\,v^2_2\over 4 \pi v^4}~,
\label{ampel}
\eeq
where $s$ is the center-of-mass energy squared.  

The scattering amplitude for $t \bar{t} \to V_L V_L$ grows in this
region since $\sqrt{s}< m_H$. The tree-level spin-zero, weak-isosinglet,
color singlet amplitude is \cite{Marciano:1989ns}
\beq
{\cal A}_{inel}(t\bar{t} \to V_L V_L)\vert_{\sqrt{s}\ll m_H}
= {\sqrt{3N_c}\, \sqrt{s}\, m_t
\over 16\pi v^2}\left(1-{\cos\alpha\over \sin\beta}\sin(\beta-\alpha)\right)
\approx {\sqrt{3N_c}\, \sqrt{s}\, m_t
\over 8\pi v^2}~.
\label{ampinel}
\eeq
The contribution proportional to $\sin(\beta-\alpha)$ in the
amplitude above arises from $h$-exchange, and while it 
is equal in magnitude to the contribution from Higgs-exchange in
the standard model it has the {\it opposite} sign!

Finally, the leading contributions to $t\bar{t}$ elastic scattering in
this energy regime comes from $Z$- and $h$-exchange.  As the coupling
of the $h$ to $t\bar{t}$ is approximately the same as that of the
standard model Higgs, {\it cf} eqn. (\ref{Httcoupling}), this amplitude
is negligible.

In order to judge the relative importance of ${\cal A}_{el}$ and ${\cal
  A}_{inel}$, we perform a coupled-channel analysis of the
spin-zero, weak-isosinglet, color-singlet $V_LV_L$ and $t\bar{t}$
states. At tree-level, the amplitudes for $t\bar{t} \to V_L V_L$ and the
reverse process $V_L V_L \to t\bar{t}$ are real and, therefore, equal.
To perform the coupled-channel analysis, we consider the
scattering matrix
\beq
\left(\matrix{{\cal A}_t\approx 0 & {\cal A}_{inel} \cr
{\cal A}_{inel} & {\cal A}_{el}} \right)~.
\eeq
Unitarity requires that the real part of the largest eigenvalue
of this matrix be less than one-half  \cite{Luscher:1988gc}.
This yields the constraint
\beq
2{\cal A}_{el} + 4 {\cal A}^2_{inel} \laem 1~.
\eeq
From this we see that the inelastic amplitude dominates the unitarity
constraints in this regime if $2{\cal A}^2_{inel} > {\cal A}_{el}$.
(Note that, the contributions to overall unitarity from {\it
  both} processes scale like $s$.)  From eqns.  (\ref{ampel}) and
(\ref{ampinel}) we see this occurs only if
\beq
v_2 \le \sqrt{3N_c \over 8 \pi} m_t \approx 105\, {\rm GeV}~.
\label{vtwoupper} 
\eeq

Can $v_2$ be this small in the two-Higgs doublet model?  Consider
scattering at high-energies, $\sqrt{s}\gg m_H$.  In this region,
$H$-exchange contributes significantly to $t\bar{t}$ elastic scattering.
The contribution to the spin-zero color singlet amplitude coming from
$H$ exchange is \cite{Chanowitz:1979mv}
\beq
{\cal A}_t(t\bar{t}\to t\bar{t})\vert_{\sqrt{s}\gg m_H} 
= -\, {N_c\, m^2_t \over 16 \pi v^2_2}~.
\eeq
Unitarity implies the absolute value of this amplitude
must be less than one-half. This yields
the bound
\beq
v_2 \ge \sqrt{N_c \over 8\pi} m_t \approx 60\, {\rm GeV}~.
\label{vtwolower}
\eeq
Comparing eqns. (\ref{vtwoupper}) and (\ref{vtwolower}), we see that one
can consistently arrange for the inelastic amplitude for $t\bar{t}\to
V_L V_L$ to dominate over the elastic amplitude for $V_L V_L \to V_L
V_L$ without violating unitarity in elastic $t\bar{t}$ scattering at
high energies.

In order to determine whether one can saturate the bound on top-quark
mass generation, however, one must see if the $H$-boson mass can be made
as large as the bound in eqn. (\ref{applchanbound}). That is, from
(\ref{Hmass}), we must ask how large $\lambda_5$ can be. Consider the
the tree-level amplitude for spin-zero $hH \to h H$ scattering. From
(\ref{lambdafiveterm}), we calculate
\beq
{\cal A}_{hH}(hH\to hH)\vert_{\sqrt{s}\gg m_H} 
= -\,{\lambda_5 \over 16\pi}~.
\eeq
Requiring that the real part of this amplitude not exceed
one half (in absolute value), we find $\lambda_5 \le 8\pi$
and hence, from (\ref{Hmass}),
\beq
m_H \le \sqrt{4\pi} v \approx\, 870\,{\rm GeV}~.
\label{mHbound}
\eeq
Since the bound (\ref{mHbound}) is much less than 3 TeV, we conclude
that it is not possible to saturate the bound on top-quark mass
generation in the two-Higgs model. This conclusion is consistent with
that obtained by considering the triviality of the model
\cite{Kominis:1993zc}.

In summary, while it is possible in the two-Higgs model to arrange for
the inelastic scattering amplitude $t\bar{t} \to V_L V_L$ to dominate
over the elastic $V_L V_L \to V_L V_L$ amplitude, one cannot saturate
the bound on top-quark mass generation while maintaining unitarity in
Higgs scattering at high energies. The situation is analogous to that of
trying to saturate the scale $\Lambda_{EW}$
\cite{Lee:1977eg,Chanowitz:1985hj,Marciano:1989ns} in the standard
one-doublet Higgs model. In that case as well, Higgs scattering
\cite{Lee:1977eg} and triviality \cite{Dashen:1983ts} preclude making
the Higgs boson as heavy as $\Lambda_{EW}$.

Finally, we note that the lower bound on $v_2$ (\ref{vtwolower}) is
approximately saturated by the ``top-Higgs'' in topcolor-assisted
technicolor models \cite{Hill:1995hp}. However, in the simplest version
of this model the top-Higgs mass is proportional to $v_2$, unlike the
relation found in eqn. (\ref{Hmass}).  It is interesting to speculate
whether the bound (\ref{applchanbound}) could be saturated in a variant
topcolor-assisted technicolor model.

\bigskip


\centerline{\bf Acknowledgments}

I thank Tom Appelquist, Bogdan Dobrescu, Howard Georgi, Tao Han,
Elizabeth Simmons, and Scott Willenbrock for discussions and comments on
the manuscript, and the Aspen Center for Physics for its hospitality
while this work was completed.  {\em This work was supported in part by
  the Department of Energy under grant DE-FG02-91ER40676.}


\bibliography{applchan}
\bibliographystyle{h-physrev}
\end{document}